\documentclass[11pt,a4paper]{article} 
\usepackage{jheppub} 

\pdfoutput=1
\makeatletter
\gdef\@fpheader{ }
\makeatother
\toccontinuoustrue

\newcommand{\E}{{\mathbf{E}}}
\newcommand{\B}{{\mathbf{B}}}
\newcommand{\J}{{\mathbf{J}}}
\newcommand{\x}{{\mathbf{x}}}
\renewcommand{\L}{{\mathcal{L}}}
\renewcommand{\H}{{\mathcal{H}}}
\newcommand{\e}{{\mathcal{E}}}
\newcommand{\q}{{\hat{q}}}
\newcommand{\p}{{\hat{p}}}
\newcommand{\F}{{\hat{F}}}
\newcommand{\C}{{\hat{C}}}

\begin{document}

\title{QED without Gauge Fields}
\author[a,b]{Mahdiyar Noorbala}
\affiliation[a]{Department of Physics, University of Tehran, Tehran, Iran.  P.O.~Box 14395-547}
\affiliation[b]{School of Astronomy, Institute for Research in Fundamental Sciences (IPM), Tehran, Iran.  P.O.~Box 19395-5531}

\abstract{We begin by studying a very simple Hamiltonian for Maxwell's equations that has no gauge fields and is made entirely of the electromagnetic fields.  We then show that this theory cannot be quantized.  We also show that no other such simple theory that only involves the electromagnetic fields can be quantized.  This gives further evidence for the important role of gauge fields in QED.}

\maketitle

\section{Introduction}

The standard variational approach to Maxwell's equations is based on the action
\begin{equation}
S[A] = \int \left( -\frac{1}{4} F_{\mu\nu} F^{\mu\nu} + J^\mu A_\mu \right) d^4x,
\end{equation}
where the metric signature is $-+++$, and $F_{\mu\nu} = \partial_\mu A_\nu - \partial_\nu A_\mu$ is the field strength tensor.  There are various physically equivalent approaches to quantization.  One of the most convenient ones is that of constrained systems \cite{Dirac}.  It begins by defining the density of momentum conjugate to $A_\mu$:
\begin{equation}
\pi^\mu := \pi_{A_\mu} = \frac{\delta S}{\delta \dot A_\mu} = F^{\mu0}, \qquad \text{or} \qquad \pi_i = -E_i.
\end{equation}
The Poisson brackets of $A_i$ and $\pi_j$ obey the standard formula for canonical variables, from which the Poisson brackets of $E_i$ and $B_j$ are to be computed:
\begin{equation}\label{stdPB}
\begin{aligned}
\{A_i(\x), A_j(\x')\} &= \{\pi_i(\x), \pi_j(\x')\} = 0, & \{A_i(\x), \pi_j(\x')\} &= \delta_{ij} \delta^3(\x-\x'), \\
\{E_i(\x), E_j(\x')\} &= \{B_i(\x), B_j(\x')\} = 0, & \{E_i(\x), B_j(\x')\} &= \{-\pi_i(\x), \epsilon_{jkl} \partial'_k A_l(\x')\}  \\
& & &= \epsilon_{ijk} \frac{\partial}{\partial x'_k} \delta^3(\x-\x').
\end{aligned}
\end{equation}
These Poisson brackets can then be promoted to commutators to quantize the theory.

If we forget about the sources for a moment and set $J=0$, we see that the action is a function of the fields $F$ alone.  Nonetheless, varying $S$ with respect to $F$ does not yield the correct equations of motion.  It seems that $A$ has a fundamental role without which Hamiltonian, action, and quantization cannot be defined.   Of course, gauge symmetry has proved to be a cornerstone of particle physics and there is no doubt that it has wonderful theoretical and experimental consequences.  But the question here is whether the quantization or at least a Hamiltonian formulation of Maxwell's equations without resort to gauge fields is possible.  Since ultimately, the electric and magnetic fields---rather than gauge fields---are the physical variables, this is a well-motivated question.

There is a conventional argument for why quantization necessitates gauge invariance \cite{Weinberg}.  It is based on the particular nature of massless spin-1 particles, and leads to the conclusion that the causal field operator made out of their creation/annihilation operators has to obey gauge symmetry.  Although this argument is compelling in many respects, it doesn't forbid construction of the Hamiltonian out of the electromagnetic field $F$.  The aim of this paper is to address this question.  But we will not take the particle theoretic point of view where the mass and spin of the particle are crucial.  Indeed, we solely concentrate on the Hamiltonian formulation of Maxwell's equations and its quantization (which we call QED).

The rest of this paper is organized as follows.  In Section \ref{sec:classic} we derive a Hamiltonian for the classical electromagnetism in which gauge fields play no role.  In Section \ref{sec:quantum} we explain the trouble with the quantization of this model and its modifications.  Finally, we summarize and conclude in Section \ref{sec:conclusions}.

\section{The Gauge-less Formulation}\label{sec:classic}

There are two of Maxwell's equations,
\begin{equation}\label{Maxwell}
\begin{aligned}
&\frac{\partial\E}{\partial t} = \nabla \times \B - \J, && &\nabla \cdot \E = \rho, \\
&\frac{\partial\B}{\partial t} = -\nabla \times \E, && &\nabla \cdot \B = 0,
\end{aligned}
\end{equation}
that contain time-derivatives and are reminiscent of Hamilton's equations, as if $\E$ and $\B$ are canonical coordinates.  They suggest that we may find a Hamiltonian, exclusively in terms of $\E$ and $\B$, whose equations of motion are Maxwell's equations.

This motivates us to try to construct a gauge-less Hamiltonian of the form $H_\text{gl}=\int \H_\text{gl}(\E,\B) d^3x$ and treat $E_i$s and $B_i$s as the generalized coordinates and momentum densities, i.e.,
\begin{equation}\label{myPB}
\{ E_i(\x), B_j(\x') \}_\text{gl} = \ell^{-1} \delta_{ij} \delta^3(\x-\x'),
\end{equation}
where $\ell$ is an irrelevant length scale required on the basis of dimensional analysis.  We note in passing that this is in sharp contrast with Eqs.~\eqref{stdPB}; we will get back to this point later.  We further demand that $\H_\text{gl}$ involve at most one spatial derivative and be quadratic in its arguments, so that the equations of motion are linear in $E_i$s and $B_i$s with first order spatial derivatives.  With these restrictions, a generic $\H_\text{gl}$ reads
\begin{equation}\label{Hansatz}
\ell^{-1} \H_\text{gl} = E_ia_{ij}E_j + B_ib_{ij}B_j + E_ic_{ij}B_j + d_iE_i + e_iB_i,
\end{equation}
where $a_{ij} = a^{(0)}_{ij} + a^{(1)}_{ijk} \partial_k$ and $a^{(0)}_{ij} = a^{(0)}_{ji}$ (similarly for $b$ and $c$), while  $d_i = d^{(0)}_i + d^{(1)}_{ij} \partial_j$ (similarly for $e$).  

To get the desired set of equations of motion in the left column of \eqref{Maxwell}, we need
\begin{equation}
\frac{\partial E_i}{\partial t} = \{E_i,H_\text{gl}\}_\text{gl}, \qquad
\frac{\partial B_i}{\partial t} = \{B_i,H_\text{gl}\}_\text{gl},
\end{equation}
which translates to:
\begin{equation}
\begin{aligned}
\epsilon_{ijk} \partial_j B_k - J_i &= b_{ij}B_j + B_jb^{(0)}_{ji} - \partial_k \left( B_jb^{(1)}_{jik} \right) + E_jc^{(0)}_{ji} - \partial_k \left( E_jc^{(1)}_{jik} \right) + e^{(0)}_i - \partial_j e^{(1)}_{ij}, \\
-\epsilon_{ijk} \partial_j E_k &= -a_{ij}E_j - E_ja^{(0)}_{ji} + \partial_k \left( E_ja^{(1)}_{jik} \right) - c_{ij}B_j - d^{(0)}_i + \partial_j d^{(1)}_{ij}.
\end{aligned}
\end{equation}
This implies that
\begin{equation}
\begin{aligned}
c_{ij}&=d_i=0, &\qquad e^{(0)}_i - \partial_j e^{(1)}_{ij} &= -J_i, \\
2b^{(0)}_{ij} &= \partial_k b^{(1)}_{jik}, & b^{(1)}_{ijk} - b^{(1)}_{jik} &= \epsilon_{ikj}, \\
2a^{(0)}_{ij} &= \partial_k a^{(1)}_{jik}, & -a^{(1)}_{ijk} + a^{(1)}_{jik} &= -\epsilon_{ikj}.
\end{aligned}
\end{equation}
So we can write $a^{(1)}_{ijk} = -\frac12 \epsilon_{ijk} + \tilde{a}^{(1)}_{ijk}$ with $\tilde{a}^{(1)}_{ijk} = \tilde{a}^{(1)}_{jik}$ (similarly for $b$); and the Hamiltonian density becomes:
\begin{equation}
\ell^{-1} \H_\text{gl} = \frac{1}{2} \E \cdot \nabla \times \E + \frac{1}{2} \B \cdot \nabla \times \B - \J \cdot \B + \frac{1}{2} \partial_k \left( \tilde{a}^{(1)}_{ijk} E_i E_j + \tilde{b}^{(1)}_{ijk} B_i B_j \right) + \partial_j \left( e^{(1)}_{ij} B_i \right).
\end{equation}
Of course, we should ignore the total derivatives---as we did so in the first place when writing Eq.~\eqref{Hansatz}---since they disappear in our final result for the Hamiltonian:\footnote{A similar expression appears, e.g., in Ref.~\cite{Escalante:2013st} and is based on the Lagrangian density ${\cal L} = -\frac14 {\bf F}^* \cdot (i\partial{\bf F}/\partial t + \nabla\times{\bf F}) + \text{c.c.}$ introduced in Ref.~\cite{Lanczos}, in which ${\bf F} = \B+i\E$.  However, these works begin by assuming $\E$ and $\B$ as independent coordinates in a $6+6$-dimensional phase space, which is different from our $3+3$-dimensional space.  Their action differs from the one we derive below in Eq.~\eqref{S}, too.}
\begin{equation}\label{Ham}
H_\text{gl} = \ell \int \left( \frac{1}{2} \E \cdot \nabla \times \E + \frac{1}{2} \B \cdot \nabla \times \B - \J \cdot \B \right) d^3x.
\end{equation}
Note that $\rho$ and $\J$ are sources that have to satisfy the continuity equation, $\partial\rho/\partial t + \nabla\cdot\J = 0$.  (For the purposes of this paper, the sources can be considered non-dynamical.  But one can associate their dynamics with electrically charged particles if those degrees of freedom are taken into account.)

So far our Hamiltonian produces six equations out of the eight Maxwell's equations.  We must supplement the remaining two Maxwell's equations (the right column of \eqref{Maxwell}) as constraints on initial conditions and ensure that they remain satisfied at all subsequent times.  This requires:
\begin{equation}
\{\partial_i E_i,H_\text{gl}\}_\text{gl} = \frac{\partial \rho}{\partial t}, \qquad \{\partial_i B_i,H_\text{gl}\}_\text{gl} = 0.
\end{equation}
But these are trivially satisfied as a consequence of the original six equations (the left column of \eqref{Maxwell}) and as a result of the continuity equation.  So once the initial conditions are chosen to satisfy the constraints, they are guaranteed to remain satisfied under the Hamiltonian evolution.  This shouldn't undermine the significance of the constraints; if it were not for them, the charge density would not appear anywhere in the solution, since $H_\text{gl}$ is independent of $\rho$.  Finally, we note that the two constraints $\nabla \cdot \E - \rho=0$ and $\nabla \cdot \B=0$ are second-class, so it is not possible to obtain a more general motion by adding them to $\H_\text{gl}$.

The reader may wonder how our Hamiltonian $H_\text{gl}$ is related to the standard Hamiltonian,\footnote{The most general Hamiltonian in the standard theory is more complicated.  Here $H_\text{std}$ governs the time evolution in the temporal gauge ($A^0=0$).}
\begin{equation}
H_\text{std} = \int \left( \frac{E^2 + B^2}{2} - \J\cdot\mathbf{A} \right) d^3x,
\end{equation}
and energy of the electromagnetic field,
\begin{equation}
\e = \frac{1}{2} \int \left( E^2 + B^2 \right) d^3x.
\end{equation}
Of course, the gauge-less formulation has to yield the standard result that when there is no dissipation, $\e$ must be a conserved quantity. 
Indeed, using the gauge-less Poisson brackets in Eq.~\eqref{myPB}, it is straightforward to show that
\begin{equation}
\frac{d \e}{dt} = \{ \e, H_\text{gl} \}_\text{gl} = -\int \left[ \nabla \cdot (\E \times \B) + \J \cdot \E \right] d^3x.
\end{equation}
Thus we have reproduced the integrated form of Poynting's theorem $d\e/dt + \int \nabla \cdot \mathbf{S} d^3x = -\int \J\cdot\E d^3x$.  In a similar manner, using the standard Poisson brackets in Eq.~\eqref{stdPB}, one obtains:
\begin{equation}
\frac{d H_\text{gl}}{dt} = \frac{\partial H_\text{gl}}{\partial t} + \{ H_\text{gl}, H_\text{std} \}_\text{std} = -\ell \int \dot\J \cdot \B d^3x.
\end{equation}
This confirms that $ \{ H_\text{gl}, H_\text{std} \}_\text{std} = 0$ as it should, since the time evolution of $H_\text{gl}$ can be obtained by either $H_\text{std}$ or by $H_\text{gl}$ itself.\footnote{Note that $H_\text{gl}$ cannot be used to generate the time evolution of $H_\text{std}$, since the latter involves gauge fields, whose Poisson brackets are undefined in the gauge-less formulation.}

Since we are going to use canonical quantization, we will not need a Lagrangian.  But to complete this section, we make a brief comment.  It is straightforward to obtain the Lagrangian density:
\begin{equation}
\L_\text{gl} = \dot\E \cdot \B - \H = \frac{1}{2} \B \cdot \nabla \times \B - \frac{1}{2} \E \cdot \nabla \times \E.
\end{equation}
Of course, $\L_\text{gl}$ has to be considered as a function of $\E$ and $\dot\E$, which can be done, using the Helmholtz theorem, by writing $\B$ in terms of $\dot\E$:
\begin{equation}\label{B(E)}
\B(x) = \nabla \times \int \frac{\dot \E(x') + \J(x')}{4\pi|\x-\x'|} d^3x' = \int \frac{(\dot \E+\J)_{x'} \times (\x-\x')}{4\pi|\x-\x'|^3} d^3x'.
\end{equation}
We also need to impose the constraint $\nabla\cdot\E=\rho$ by a Lagrange multiplier $u(x)$ in the Lagrangian (there is no need to impose $\nabla \cdot \B=0$, since it is already satisfied by Eq.~\eqref{B(E)}---in the Lagrangian formalism $\B$ is \textit{defined} by Eq.~\eqref{B(E)}).  So finally the action corresponding to $H_\text{gl}$ reads:
\begin{equation}\label{S}
\begin{split}
S_\text{gl}[\E] = \frac{1}{8\pi} \int \frac{\x-\x'}{|\x-\x'|^3} \cdot \left[ (\dot\E+\J)_x \times (\dot\E+\J)_{x'} \right] d^3x' d^4x \\
- \frac{1}{2} \int (\E \cdot \nabla \times \E)_x d^4x + \int u(x) (\nabla\cdot\E-\rho)_x d^4x.
\end{split}
\end{equation}

It is easy to see that $\delta S_\text{gl}=0$ implies $\nabla^2 u=0$.  If the fields vanish at infinity, the constraint is satisfied and we can have $u=0$ there.  $\nabla^2 u=0$ then implies that $u=0$ elsewhere.  Thus we can set $u=0$ in all equations of motion (after performing the variation) to obtain $\partial \B/\partial t = -\nabla \times \E$.  Gauss's law is imposed as a constraint, and the remaining two of Maxwell's equations follow from the definition of the conjugate momentum in the action.  Note how this differs from the standard formulation in which the sourced equations follow from the action, while the sourceless ones are automatic.  Also note that we have suppressed the time dependence in Eqs.~\eqref{B(E)} and \eqref{S}: all quantities are evaluated at the same time $x^0=x'^0$.  Therefore, we observe that although the action is local in time, it is very non-local in space.  In contrast, the Hamiltonian $H_\text{gl}$ is local in both time and space.\footnote{By ``local in space'' we mean the standard terminology that the second and higher spatial derivatives of the fields do not appear in $\H$.}

There is another aspect of the gauge-less formulation which is more important than the non-locality of action, and that is the lack of Lorentz invariance.  In particular, unlike the standard Poisson brackets in Eq.~\eqref{stdPB}, our Poisson brackets in Eq.~\eqref{myPB} are not Lorentz invariant.  Despite this, the gauge-less formulation leads to Maxwell's equations which are clearly Lorentz invariant.  As long as we are only concerned about the equations of motion, none of these problems is an issue.  However, as we see in the next section, Lorentz violation does obstruct the quantization process of the gauge-less formulation.

\section{Quantization}\label{sec:quantum}

As we remarked before, the fact that the Poisson brackets in Eq.~\eqref{myPB} are not covariant is not an issue in the classical theory, since all that matters there is the equation of motion.  When we quantize the theory, however, the commutator of two observables has physical meaning; it is related to the uncertainty after all.  Let us inspect in more details this distinction between the classical and quantum theory.

Quite generally, let $q^a$ and $p_a$ be the generalized coordinates and momenta describing a given system in one reference frame, and let $q'^a$ and $p'_a$ describe the same system in a second frame.  Of course, we must have symplectic structures $\{,\}$ and $\{,\}'$ on both sets of coordinates satisfying $\{q^a,p_b\}=\delta^a_b$ and $\{q'^a,p'_b\}'=\delta^a_b$.  Furthermore, the primed coordinates are related to the unprimed ones by Lorentz transformations and we write $q'(q,p)$ and $p'(q,p)$ to express this relationship.  In most applications, the primed and unprimed Poisson brackets are the same, so that $\{q'^a,p'_b\}' = \{q'^a(q,p),p'_b(q,p)\}$.  But this is not a necessity in a classical theory.  Our formulation of Maxwell's equations in the previous section is one such example, where $\{E'_i,B'_j\}' \neq \{E'_i(E,B),B'_j(E,B)\}$.  To be specific, under a boost in the $x$-direction, we have $E'_2 = \gamma(E_2-vB_3)$ and $E'_3 = \gamma(E_3+vB_2)$, which implies $\{E'_2(\x),E'_3({\bf y})\} = 2v\gamma^2 \delta^3(\x-{\bf y}) \neq 0$.  Nevertheless, the gauge-less description can successfully reproduce Maxwell's equations.

The commutator of two observables is not as flexible as their Poisson bracket.  The reason lies in the fact that probability is invariant and so the Hilbert spaces of the two observers must be unitarily related.  In practice, it is common to use a single Hilbert space for both observers and employ identical operators to describe physical observables.  Then we relate the state vector as seen by one observer by a unitary transformation to the state vector as seen by the other observer.  Thus, there is only one kind of commutator; symbolically: $[,]=[,]'$.  In particular, for the canonical operators $\q^a$ and $\p_b$, we must have $[\q'^a(\q, \p), \p'_b(\q, \p)] = [\q^a, \p_b]$.

We are thus led to conclude that the gauge-less formulation of the previous section cannot be quantized since its Poisson brackets are not covariant under Lorentz transformations.  The natural question is whether it is possible to write down a set of covariant Poisson brackets involving the physical fields (rather than gauge fields).  We now try to answer this question under the assumption that the canonical coordinates of interest are linear combinations of the electric and magnetic fields, to allow for the pairing of coordinates and their conjugate momenta to be different from that of Eq.~\eqref{myPB}.  

Specifically, let us assume that there are three pairs of canonically conjugate fields, collectively denoted by $\q(x)$s and $\p_\mu(x)$s, which are linear combinations of the six nonzero components of $\F_{\mu\nu}(x)$.\footnote{Useful properties of commutators of quantum fields in the notation appropriate for this discussion are summarized in Appendix~\ref{app-qft}.}  The $\q$s and $\p_\mu$s must be organized as elements of some tensor in order for the commutators to be covariant.  Therefore, the $\q$s and $\p_\mu$s are linearly related to $\F_{\mu\nu}$s with coefficients that are invertible tensors.  We take the coefficients appearing in these linear combinations to be independent of $x$, to respect translation invariance.  Inverting these relations, we can write $\F_{\mu\nu}$s as linear combinations of $\q$s and $\p$s.  Since the commutators of the latter are given by Eq.~\eqref{covcom}, the commutator of the former must be of the form:
\begin{equation}\label{com-FF}
[ \F_{\mu\nu}(x), \F_{\rho\sigma}(0) ] = i{\hat A}_{\mu\nu\rho\sigma}(x) \Delta(x) + i{\hat B}_{\mu\nu\rho\sigma}^\tau(x) \partial_\tau \Delta(x).
\end{equation}
The operator $\hat B$ cannot be identically zero, since $[\q,\p]$s are linear in $[\F,\F]$s after all.  In fact, by virtue of the constraint on the coefficient of $\partial_\mu \Delta$ in Eq.~\eqref{covcom}, $\hat B$ must remain a nonzero tensor as $x$ approaches the origin.  Furthermore, this nonzero tensor cannot depend on the direction of $x$ (for example, it cannot be $x^\mu x^\nu/x^2$), because we took the coefficients relating $\q$s and $\p_\mu$s to $\F_{\mu\nu}$s to be $x$-independent.  Altogether, these imply that the tensorial structure of ${\hat B}_{\mu\nu\rho\sigma}^\tau$ must be made up entirely of the invariant tensors, namely, $\eta_{\mu\nu}$ and $\epsilon_{\mu\nu\rho\sigma}$.  This brings about a contradiction, since $\hat B$ has an odd number of indices and cannot be possibly made up of the even-ranked tensors $\eta$ and $\epsilon$.\footnote{We have also assumed that there is no preferred 4-vector in the theory.  We can turn off the sources $J^\mu$ for the sake of this discussion, since if---as we showed---the theory doesn't work in the absence of the sources, it won't work in their presence either.} 

We conclude that the electric and magnetic fields themselves (or linear combinations of them) cannot provide a Lorentz invariant set of canonical variables.  Note how the standard formulation avoids this conclusion: Since the electromagnetic fields are derivatives of the canonical variables (i.e., the gauge fields), their commutator contains two derivatives of $\Delta$ (equivalently, one derivative of $\delta^3$) and $\hat B$ is identically zero.  There is no obstruction in constructing the tensor ${\hat A}_{\mu\nu\rho\sigma}$ in Eq.~\eqref{com-FF} out of the metric and $x^\mu$ (equivalently $\partial_\mu \Delta$).  Indeed, for the free Maxwell theory, we have 
$[ \F_{\mu\nu}(x), \F_{\rho\sigma}(x') ] = i \partial_{[\mu} \eta_{\nu][\sigma} {\partial_{\rho]}}' \Delta(x-x')$.

\section{Conclusions}\label{sec:conclusions}

We saw that a Hamiltonian formulation of Maxwell's equations without resort to gauge fields is possible.  This gauge-less classical theory has a local Hamiltonian, but its action is non-local in space.  Although, the theory ultimately yields Lorentz invariant equations of motion, the action and the canonical structure are not preserved under Lorentz transformations.  This is not a problem in a classical theory, but it does affect the quantum theory.  

We then showed that the gauge-less model cannot be modified to have covariant commutators.  We did so by considering all possible ways of writing three canonical variables and their conjugate momenta as linear combinations of the electric and magnetic fields.  It turns out that this is inconsistent with what we expect from the commutators of electromagnetic fields.  Of course, more complicated relationships between the canonical variables and the electromagnetic fields can be studied.  But we restricted our attention to the simplest possibility where the electromagnetic fields themselves are the building blocks of the canonical formalism.

\section{Acknowledgements}
I acknowledge financial support from the research council of University of Tehran.

\appendix

\section{Commutators of Quantum Fields}\label{app-qft}
In this appendix we review some basic properties of commutators of generic (free and non-free) relativistic quantum fields.  Since we have undertaken a Hamiltonian approach, we use the notation $\q$ and $\p$ for canonical variables even when they are field operators.

Let $n=(1,{\bf 0})$ be the 4-velocity of an observer, and $\x$ be the proper spatial coordinates on the surface $t=0$.  According to this observer, the canonical fields obey
\begin{equation}\label{ncom-qp}
[ \q_n(\x),\q_n(\x') ] = [ \p_n(\x),\p_n(\x') ] = 0, \qquad [ \q_n(\x),\p_n(\x') ] = i\delta^3(\x-\x') \hat1.
\end{equation}
For a scalar field, the Heisenberg field operator $\q(x)$ is obtained by evolving $\q_n(\x)$ with the Hamiltonian $H_n(\q_n,\p_n)$ to any future or past time.  It encapsulates all of the information in $\q_n(\x)$s for all timelike $n$ and all spacelike $\x$.  In particular, $[ \q(x),\q(0) ] = 0$ for spacelike $x$.  For timelike $x$, this provides even more information and we have
\begin{equation}
[\q(x),\q(0)] = i\C^{(qq)}(x) \Delta(x),
\end{equation}
where the operator $\C^{(qq)}$ contains this extra information (for a free field of mass $m$, it is equal to the identity operator).
\begin{equation}
\Delta(x) = \frac{i}{(2\pi)^3} \int \operatorname{sgn}(k^0) \delta(k^2+m^2) e^{ik_\mu x^\mu} d^4k
\end{equation}
is Schwinger's $\Delta$ function with the following properties: (i) $\Delta(-x) = -\Delta(x)$, so that $\Delta(0)=0$; (ii) $\Delta(x)$ vanishes for spacelike $x$; and (iii) for any spacelike surface $\Sigma$ passing through the origin and having normal vector $\sigma^\mu$:
\begin{equation}\label{intDelta}
\int_\Sigma \partial_\mu \Delta(x) d\sigma^\mu = -1.
\end{equation}

Unlike the $\q_n(\x)$s which give rise to the $n$-independent $\q(x)$, the $\p_n(\x)$ do not combine into a scalar operator $\p(x)$, rather into a 4-vector $\p_\mu(x)$ such that
\begin{equation}
[\q(x),\p_\mu(0)] = -i\C^{(qp)}(x) \partial_\mu\Delta(x), \qquad \C^{(qp)}(0) = \hat1.
\end{equation}
Using Eq.~\eqref{intDelta}, the desired commutation relations in \eqref{ncom-qp} are obtained for $\p_n=n^\mu \p_\mu$.  Again the operator $\C^{(qp)}$ carries additional information (for a free field, $\p_\mu = \partial_\mu \q$ and $\C^{(qp)} = \hat1$).

The commutator of the momenta at spacelike separation must vanish, too.  So we have 
\begin{equation}
[\p_\mu(x),\p_\nu(0)] = -i\C^{(pp)}_{\mu\nu}(x) \Delta(x).
\end{equation}
For a free field and timelike $x$, $\C^{(pp)}_{\mu\nu} = \hat1 (\partial_\mu \partial_\nu \Delta) / \Delta$.

The above discussion can be extended to the case where $\q^a$ is a collection of fields labeled by $a$ (which could be spacetime indices), each conjugate to $\p_{a\mu}$.  The obvious modification is:
\begin{equation} 
\begin{aligned}\label{covcom}
{}[\q^a(x),\q^b(0)] &= i \C^{(qq)ab}(x) \Delta(x), \\
[\q^a(x),\p_{b\mu}(0)] &= -i\C^{(qp)a}_b(x) \partial_\mu\Delta(x), \qquad \C^{(qp)a}_b(0) = \delta^a_b \hat1, \\
[\p_{a\mu}(x),\p_{b\nu}(0)] &= -i\C^{(pp)}_{ab\mu\nu}(x) \Delta(x).
\end{aligned}
\end{equation}
where $\delta$ represents a product of Kronecker deltas.


\begin{thebibliography}{9}

\bibitem{Dirac}
  P.A.M.~Dirac,
  \textit{Lectures on Quantum Mechanics},
  Yeshiva University, New York, U.S.A.\ (1964).

\bibitem{Weinberg} 
  S.~Weinberg,
  \textit{The Quantum Theory of Fields, Volume I: Foundations},
  Cambridge University Press, Cambridge, U.K.\ (1995).
  
\bibitem{Escalante:2013st} 
  A.~Escalante and O.~R.~Tzompantzi,
  \textit{Hamiltonian Dynamics for an alternative action describing Maxwell's equations},
  \textit{Int.\ J.\ Pure Appl.\ Math.\ } \textbf{81} (2012) 701
  [arXiv:1301.0502 [math-ph]].

\bibitem{Lanczos}
  C.~Lanczos, 
  \textit{The Variational Principles of Mechanics}, 4th ed.,
  University of Toronto Press, Toronto, Canada (1970).
  
\end{thebibliography}
\end{document}